\documentclass[sn-nature]{sn-jnl}

\usepackage{graphicx}%
\usepackage{multirow}%
\usepackage{amsmath,amssymb,amsfonts}%
\usepackage{amsthm}%
\usepackage[title]{appendix}%
\usepackage{xcolor}%
\usepackage{textcomp}%
\usepackage{manyfoot}%
\usepackage{booktabs}%
\usepackage{algorithm}%
\usepackage{algorithmicx}%
\usepackage{algpseudocode}%
\usepackage{listings}%
\usepackage{siunitx}

\begin{document}

\title[Article Title]{Astrophysical aspects of $^{12}\mathrm C(p, \gamma)^{13}\mathrm N$ reaction}


\author[1,2]{\fnm{Soumya} \sur{Saha}}
\email{ssphy.caluniv@gmail.com}



\affil[1]{\orgdiv{Department of Physics}, \orgname{Bidhan Chandra College, Rishra}, \orgaddress{\street{31, G.T Road (East)}, \city{Hooghly}, \postcode{712248}, \state{West Bengal}, \country{India}}}

\affil[2]{\orgdiv{Department of Physics}, \orgname{University of Calcutta}, \orgaddress{\street{92, Acharya Prafulla Chandra Road}, \city{Kolkata}, \postcode{700009}, \state{West Bengal}, \country{India}}}



\abstract{The Carbon-Nitrogen-Oxygen (CNO) cycle is fundamental to the process of hydrogen burning in stars, serving as a pivotal mechanism. At its core, the primary reaction involves the radiative capture of a proton by $^{12}\mathrm C$, crucially influencing the isotopic ratio of $^{12}\mathrm C$ to $^{13}\mathrm C$ observed in celestial bodies, including our Solar System. We have addressed this reaction mechanism by extrapolating to low-energy cross sections and S-factors with the aid of astrophysical R-matrix. Our investigation aims to shed light on its implications for nuclear reaction rates, thus influencing the abundance ratio of $^{12}\mathrm C$ to $^{13}\mathrm C$ in the cosmic environment.}




\maketitle

\section{Introduction}\label{sec1}
The ratio of $^{12}\mathrm C$/$^{13}\mathrm C$ serves as a minute indicator of the extent of stellar nucleosynthesis \cite{ss}, making it invaluable as a tracer for galactic chemical evolution. However, the fluctuations in the isotopic ratio of carbon following the dredge-up phenomena in Asymptotic Giant Branch (AGB) stars heavily rely on the chosen rate for the radiative proton capture reaction on $^{12}\mathrm C$ \cite{GLT}. In helium burning, $^{12}\mathrm{C}$ is primarily produced through the triple-alpha process, wherein three $^{4}\mathrm{He}$ fuse to form a carbon-12 nucleus. However, the production of $^{13}\mathrm{C}$ involves a different pathway. $^{13}\mathrm{C}$ is predominantly synthesized through the $^{12}\mathrm{C}(p, \gamma)^{13}\mathrm{N}$ reaction, resulting in the formation of an excited state of $^{13}\mathrm{N}$. This excited state then undergoes a $\beta^+$ decay, transforming into $^{13}\mathrm{C}$. 
The Gamow peak for the $^{12}\mathrm C (p, \gamma) ^{13}\mathrm N$ reaction in typical AGB hydrogen
burning shells occur at temperature of 0.1 GK. For AGB stars, the Direct Capture (DC) component primarily contributes since the Gamow peak window falls below the first resonance of $^{13}\mathrm N$. Its significance is further emphasized for smaller stars, where the Gamow peak window resides in an even lower energy region. Consequently, precise values
 of the cross section or S-factor at low energies become essential for making reliable predictions regarding the evolution of $^{12}\mathrm C$ abundances within stellar interiors.
J. Vogl \cite{Vogl} investigated the $^{12}\mathrm C (p, \gamma) ^{13}\mathrm N$ reaction as part of his Ph.D. thesis. The energy window covered in the measurement ranged from 150 keV to 680 keV. However, the precision of the cross-section at the lowest energies was found to be poor, with measurement errors ranging from 18\% to 92\% below 228 keV.
Rolfs et al. \cite{Rolfs} analyzed the $^{12}\mathrm C (p, \gamma) ^{13}\mathrm N$ reaction. The experiment was employed on the two $^{13}\mathrm N$  excited states at 2365 keV and at 3512 keV respectively. The differential cross sections were measured at $\theta = 0^\circ$ and $\theta = 90^\circ$. The cross-section measurement was also extended to proton energies as low as 150 keV. The extrapolation of the S-factor yielded results in good agreement with previous studies, with a precision of 14\%. Recently, we obtained an extensive dataset covering proton energies $E_p =$300 keV to 1900 keV \cite{GLT}, facilitating extrapolation into low-energy cross sections and S-factors for this significant reaction. In this regard, we utilize the R-matrix method with the AZURE2 code\cite{Azuma, AZURE2}. Interested readers are encouraged to refer to our recent publication \cite{SG} employing the R-matrix approach for a more detailed understanding of the methods applied here.

\section{Calculation}\label{sec2}
We have reanalyzed the astrophysical S-factor(or cross section), for the $^{12}\mathrm C (p, \gamma) ^{13}\mathrm N$ reaction at energies below 2.0 MeV. This analysis incorporates the contributions from two resonances with energies $E(r)_1$ = 421 keV $(J^\pi = 1/2+$) and $E(r)_2$ = 1556 keV $(J^\pi = 3/2-$), as well as the direct capture process and their interference effects.
The direct capture of a proton by a $^{12}\mathrm C$ nucleus in the $^{12}\mathrm C (p, \gamma) ^{13}\mathrm N$ reaction predominantly occurs via two specific single-particle transitions: E1 transitions from the $s_{1/2}$ state to the $p_{1/2}$ state, and from the $d_{3/2}$ state to the $p_{1/2}$ state\cite{Nesaraja}. These transitions involve the excitation or de-excitation of individual nucleons, leading to the formation of $^{13}\mathrm N$ in its ground state. 
The ground state of $^{13}\mathrm N$, characterized by a nuclear spin $(J )$ of $1/2-$, is elucidated as a scenario where a proton with a total angular momentum $( j )$ of $p_{1/2}$ is coupled to the $^{12}\mathrm C$ core, which possesses an intrinsic spin of $0+$\cite{Huang}. 
In the R-matrix framework, the absolute normalization of the direct capture amplitude in the $^{12}\mathrm C (p, \gamma) ^{13}\mathrm N$ reaction is typically expressed using previously indirectly measured ANC values\cite{Burtebaev}. The significant Coulomb repulsion between the colliding particles and the relatively low binding energy of $^{13}\mathrm N$  in the $(p + ^{12}\mathrm C)$ channel (with a Q value of 1.9435 MeV) implies the peripheral nature of the direct radiative capture process in the $^{12}\mathrm C (p, \gamma) ^{13}\mathrm N$ reaction\cite{Rolfs}. The radial component of the overlap function for the bound state wave functions of $^{12}\mathrm C$ and $^{13}\mathrm N$ in the direct capture amplitude can be approximated by the overall normalization of the direct capture part of the cross section, or the astrophysical S-factor. This approach enables the expression of the cross section or astrophysical S-factor of the $^{12}\mathrm C (p, \gamma) ^{13}\mathrm N$ reaction in terms of the single-particle (proton) ANC (Asymptotic Normalization Coefficient). The ANC determines the amplitude of the tail of the single-particle radial wave function for the bound state of $^{13}\mathrm N$, formed from the interaction between a proton and a  $^{12}\mathrm C$ nucleus.
Enhanced incorporation of the Asymptotic Normalization Coefficient (ANC), or its equivalent, the Nuclear Vertex Constant (NVC)\cite{Burtebaev}, within the R-matrix method significantly diminishes uncertainties inherent in calculating the direct capture (external) component of the amplitude for the $^{12}\mathrm C (p, \gamma) ^{13}\mathrm N$ reaction. This improvement is primarily achieved by utilizing hard-sphere scattering phase shifts to accurately model p$^{12}\mathrm C$-scattering in the initial state. This refined approach facilitates a more precise determination of the direct capture amplitude, thereby leading to a reduction in associated uncertainties.
By fitting experimental data to minimize $\chi^2$, the resonance amplitude can be accurately determined. In this scenario, the astrophysical S-factor for the $^{12}\mathrm C (p, \gamma) ^{13}\mathrm N$ reaction at relatively low energies serves as an independent source of supplementary information regarding the values of the proton and gamma widths for the resonance states in $^{13}$N.
The table presented (c.f Table \ref{tab0}) contains the optimal R-matrix parameters derived from fitting for the compound nucleus $^{13}$N, along with the permissible decay channels. Aside from the decay channels listed in Table \ref{tab0}, other narrower excited states within the same energy range were excluded due to their negligible contribution to the studied energy range. 
The value of $\Gamma^{\gamma}_1$ reported by various authors falls within the range of 0.45 to 0.67 eV\cite{Burtebaev}. Consequently, in our analysis, we adjust the width parameters of $\Gamma^{\gamma}_1$ and $\Gamma^{p}_1$, the resonant energies, and the channel radius $( r_c $) through fitting to the experimental data from Ref.\cite{GLT}. This fitting process aims to minimize $\chi^2$ exclusively within the energy range of astrophysical importance.
We have permitted a maximum orbital momentum of 4, a maximum gamma multipolarity of 3, and a maximum of 2 gamma multiplicities per decay. Our calculation does not include intermediate transitions that lead to the ground state.

\begin{table}[h]
\centering
\caption{Best-fit R-matrix parameters in the compound nucleus $^{13}$N and the allowed decay channels.}
\label{tab0}
\begin{tabular}{ccccc}
\toprule
Level & Energy & Channel & $l/$ & Fitted \\
Spin & [MeV] & Pair & Multi- & parameters \\
~ & ~ & ~ & polarity & ~ \\
\midrule
$1/2-$ & $0.0$ & $^{12}\mathrm{C}+p$ & $1$ & ANC= $1.525$ fm$^{-1/2}$ \\
\midrule
$1/2+$ & $2.375$ & $^{12}\mathrm{C}+p$ & $0$ & $\Gamma^p=35.87$ keV \\
~ & ~ & $^{13}\mathrm{N}+\gamma$ & $E1$ & $\Gamma^\gamma = 519.37$ meV \\
\midrule
$3/2-$ & $3.502$ & $^{12}\mathrm{C}+p$ & $1$ & $\Gamma^p =44.95$ keV \\
~ & ~ & $^{13}\mathrm{N}+\gamma$ & $M1$ & $\Gamma^\gamma= 542.03$ meV \\
\midrule
$5/2+$ & $3.547$ & $^{12}\mathrm{C}+p$ & $2$ & $\Gamma^p= 47.00$ keV \\
~ & ~ & $^{13}\mathrm{N}+\gamma$ & $M2$ & $\Gamma^\gamma= 0.01$ meV \\
\midrule
$1/2+$ & $14.40$ & $^{12}\mathrm{C}+p$ & $0$ & $\Gamma^p = 82.47$ eV \\
~ & ~ & $^{13}\mathrm{N}+\gamma$ & $E1$ & $\Gamma^\gamma= 0.03$ meV \\
\bottomrule
\end{tabular}
\end{table}

\section{Results}\label{sec3}
The Direct Capture (DC) model\cite{Nesaraja}characterizes a reaction involving two structureless particles: the projectile and the target. Within this framework, radiative capture is treated as a one-step process devoid of intermediate or compound nuclear states. This reaction proceeds as a transition from the initial continuum state to the final bound state under the influence of the electromagnetic interaction.
It is noteworthy that the solid line in our current study (c.f, Fig. \ref{fig1} or, Fig. \ref{fig2}), which incorporates DC along with the first and second resonances, closely matches the experimental cross-sectional data. This alignment is likely attributed to the inclusion of the second resonance, as neglecting it leads to a notable deviation. Particularly, the deviation becomes more pronounced at the higher energy tail of the first resonance in the $^{12}\mathrm C (p, \gamma) ^{13}\mathrm N$ reaction unless we include second resonance in our study. Notably, when all resonances are omitted, pure DC calculations fail to accurately predict the absolute cross sections.
\subsection{Reaction cross section or S-factor}
\begin{figure}[hhtb]
\centerline{\includegraphics[scale=0.90]{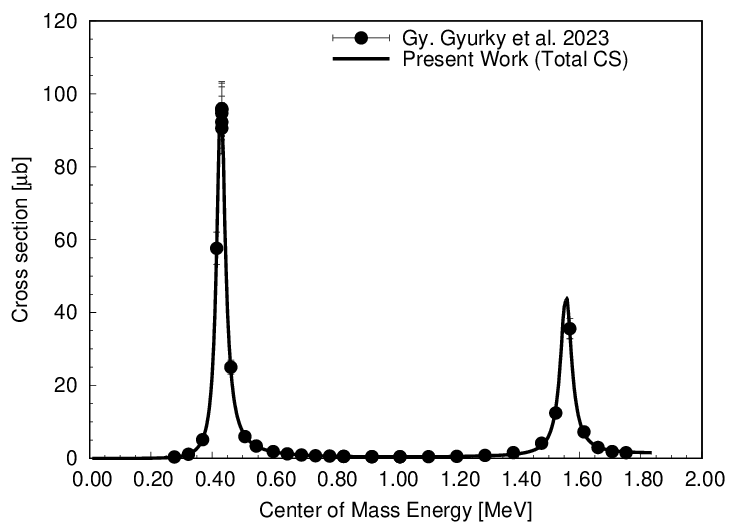}}
\caption{\label{fig1} Variation of $^{13}\mathrm N$ total reaction cross section with CM energy for $ ^{12}\mathrm C(p,\gamma)^{13}\mathrm N$ using best-fit R-matrix analysis. Plotted data with errorbars are extracted from Gy. Gy\"urky et al. 2023\cite{GLT}.}
\end{figure}

\begin{figure}
\centerline{\includegraphics[scale=0.90]{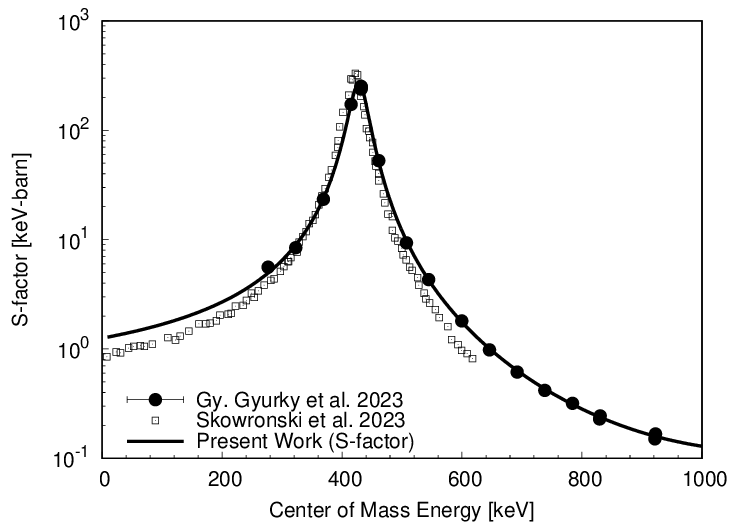}}
\caption{\label{fig2} Variation of $^{13}\mathrm N$ captured state astrophysical S-factor with CM energy for $ ^{12}\mathrm C(p,\gamma)^{13}\mathrm N$ using best-fit R-matrix analysis. Plotted data with errorbars are taken from Gy. Gy\"urky et al. 2023\cite{GLT}. The R-matrix plot from Skowronski et al.\cite{Skowronski}(retrieved using GSYS2.4\cite{Gsys}) is also shown therein.}
\end{figure}

Fig. \ref{fig1} illustrates the cross section and Fig. \ref{fig2} depicts the astrophysical S-factor for the $\gamma$-ray transition to the ground state in $^{13}$N, both as functions of the CM Energy. The resulting $\chi^2$ per degrees of freedom is approximately 1.01. In our work, the reduced-width amplitude is obtained by summing the internal (real) and external (or channel) reduced width amplitudes\cite{Burtebaev}. In addition, the R-matrix plot from Skowronski et al. \cite{Skowronski}, obtained using GSYS2.4 \cite{Gsys}, is also depicted in the Fig. \ref{fig2}.

\subsection{Thermonuclear reaction rates}
 We use nuclear reaction rate $N_A <\sigma v>$ in 
$cm^3 mol^{-1}s^{-1}$, the energies $E$ and $k_B T$ in MeV, and the cross section $\sigma$ in barn, the reaction rate becomes 

$N_A < \sigma v>$ = $3.7313 \times 10^{10} A^{-1/2} T_9 ^{-3/2} \int_0^\infty \sigma E exp(-11.605E/T_9) dE$;
Where $A$ is the reduced mass in a.m.u, and $T_9$ is the temperature in units of $10^9 K$.

The cross section can be presented as $\sigma(E) = S(E) exp(-2\pi \eta) \dfrac{1}{E}$, where $S(E)$ refers to the astrophysical S-factor. The quantity $\eta = \dfrac {Z_1 Z_2 e^2}{\hbar v}$ is the Sommerfeld parameter, $Z_1$ and $Z_2$ being the interacting nuclei charge numbers, and $\hbar$ the reduced Planck constant.

The reaction rates ($N_A < \sigma v>$) obtain by R-matrix analysis within the stellar temperature range of $0.1\leq T_9\leq 1.0$ are displayed in Table \ref{tab1}. Indeed numerical integration of an excitation curve is restricted to broad resonance structures. If the resonances being studied become excessively narrow, numerical integration may prove unreliable. Therefore, we exercise caution when the total width of the level is less than approximately 1 keV during the operation. When dealing with narrow resonances, the contributions to the reaction rate are frequently computed by summing up the individual narrow level approximations. Roughton et al. \cite{Roughton} computed nuclear reaction rates starting from a temperature of 1.0 GK, and later references \cite{Burtebaev} compared their results with the work of Ref\cite{Roughton}. However, our results essentially focus on the lower energy region of astrophysical importance. In Table. \ref{tab1} the comparison involves the reaction rates of the $^{12}\mathrm{C}(p,\gamma)^{13}\mathrm{N}$ nuclear reaction derived from our work, which employed R-matrix analysis. We juxtapose these results with those obtained by CF88\cite{CF88} utilizing an analytical approach.
\begin{table}[h]
\centering
\caption{The reaction rates of the $^{12}\mathrm{C}(p,\gamma)^{13}\mathrm{N}$ nuclear reaction resulted from this work (R-matrix analysis) are compared with CF88 \cite{CF88} (analytical approach)}
\label{tab1}
\begin{tabular}{ccc}
\toprule
Temperature & \multicolumn{2}{c}{Reaction Rates in $cm^3 mol^{-1}s^{-1}$} \\
\cmidrule{2-3}
in $GK$ & This Work & CF88 \\
\midrule
                0.10      &     1.75E-05 & 2.07E-05\\
                0.15      &     6.31E-04& 7.40E-04\\
                0.20       &   6.40E-03&7.18E-03\\
                0.25       &   3.70E-02&3.87E-02\\
                0.30       &    0.170&1.75E-01\\
                0.35       &    0.674&7.46E-01\\
                0.40       &     2.226&2.70E+00\\
                0.45       &     6.057&7.89E+00\\
                0.50       &     13.871&1.90E+01\\
                0.55      &     27.580&--\\
                0.60       &     48.974&7.10E+01\\
                0.65       &     79.459&--\\
                0.70       &     119.91&1.79E+02\\
                0.75       &     170.641&--\\
                0.80       &     231.456&3.52E+02\\
                0.85       &     301.742&--\\
                0.90       &     380.581&5.86E+02\\
                0.95       &     466.859&--\\
                1.00       &     559.358&8.68E+02\\
\bottomrule
\end{tabular}
\end{table}
The adoption rates within NACRE\cite{nacre} closely mirror those outlined in CF88\cite{CF88}, save for the divergence observed around $T_9$ = 0.30. This variance likely stems from the disparate methodologies employed: NACRE relies on numerical methods, while CF88 leans on analytical approximations. Nonetheless, our research aligns closely with the reaction rates posited by CF88, particularly at lower temperatures, notably at $T_9$ = 0.30, with an uncertainty margin of less than 2\%. This alignment underscores the significance of our present R-matrix analysis.

\section{Conclusion}
We have reanalyzed the total cross-section or the astrophysical S-factor for the $^{12}\mathrm{C}(p,\gamma)^{13}\mathrm{N}$ reaction, which is of current interest. A fresh estimation of resonance $\gamma$-width and ANC yields a well-fitted cross-section and astrophysical S-factor for the investigated reaction. This estimation effectively populates the first and second excited states of $^{13}\mathrm{N}$ and minimizes dependence on parameters within the R-matrix approach.
The nuclear reaction rates are also provided herein. It is evident that the reaction rate at 1.0 GK is somewhat less pronounced than that reported by Roughton et al. \cite{Roughton}. Given our primary focus on low-energy nuclear astrophysics, both the abundance and yield of reactions in stellar environments have direct consequences on the nucleosynthesis process, particularly as it proceeds via the CNO cycle.

\section*{Acknowledgements} 
I express my gratitude to my guru Prof. G. Gangopadhyay for his valuable suggestions. 



\bibliography{sn-bibliography}

\end{document}